\documentclass[journal=jacsat,manuscript=article]{achemso}

\usepackage{chemformula} 
\usepackage[T1]{fontenc} 
\mciteErrorOnUnknownfalse

\author{Samuel P. Gleason}
\affiliation{Department of Chemistry, University of California, Berkeley, California 94720, United States
}
\alsoaffiliation
{Materials Sciences Division, Lawrence Berkeley National Laboratory, Berkeley, California 94720, United States}
\email{smglsn12@berkeley.edu,
paul.alivisatos@uchicago.edu }
\author{Jakob C. Dahl}
\affiliation{Department of Chemistry, University of California, Berkeley, California 94720, United States
}
\alsoaffiliation
{Materials Sciences Division, Lawrence Berkeley National Laboratory, Berkeley, California 94720, United States}
\alsoaffiliation
{Molecular Foundry, Lawrence Berkeley National Laboratory, Berkeley, California 94720, United States}
\author{Mahmoud Elzouka}
\affiliation{Lawrence Berkeley National Laboratory, Berkeley, CA, United States}
\author{Xingzhi Wang}
\affiliation{Department of Chemistry, University of California, Berkeley, California 94720, United States
}
\alsoaffiliation
{Materials Sciences Division, Lawrence Berkeley National Laboratory, Berkeley, California 94720, United States}
\author{Dana O. Byrne}
\affiliation{Department of Chemistry, University of California, Berkeley, California 94720, United States
}
\alsoaffiliation
{Materials Sciences Division, Lawrence Berkeley National Laboratory, Berkeley, California 94720, United States}
\author{Mumtaz Gababa}
\affiliation{Department of Chemistry, University of California, Berkeley, California 94720, United States
}
\author{Hannah Cho}
\affiliation{Department of Chemistry, University of California, Berkeley, California 94720, United States
}
\author{Ravi Prasher}
\affiliation{Lawrence Berkeley National Laboratory, Berkeley, CA, United States}
\author{Sean Lubner}
\affiliation{Lawrence Berkeley National Laboratory, Berkeley, CA, United States}
\author{Emory Chan}
\affiliation
{Molecular Foundry, Lawrence Berkeley National Laboratory, Berkeley, California 94720, United States}
\author{A. Paul Alivisatos}

\affiliation{Department of Chemistry, University of California, Berkeley, California 94720, United States
}
\alsoaffiliation
{Department of Materials Science and Engineering, University of California, Berkeley, California 94720, United States
}
\alsoaffiliation
{Materials Sciences Division, Lawrence Berkeley National Laboratory, Berkeley, California 94720, United States}
\alsoaffiliation
{Kavli Energy NanoScience Institute, Berkeley, California 94720, United States}
\alsoaffiliation
{Current Affiliation: Department of Chemistry, University of Chicago, Chicago, Illinois 60637, United States}

\title[An \textsf{achemso} demo]
  {AuNR-SMA: Automated Gold Nanorod Spectral Morphology Analysis Pipeline}

\abbreviations{UV-VIS,AuNR,AuNRs}

\begin{document}


\begin{abstract}
The development of a colloidal synthesis procedure to produce nanomaterials of a specific size with high shape and size purity is often a time consuming, iterative process. This is often due to the time, resource and expertise intensive characterization methods required for quantitative determination of nanomaterial size and shape. Absorption spectroscopy is often the easiest method of colloidal nanomaterial characterization, however, due to the lack of a reliable method to extract nanoparticle shapes from absorption spectroscopy, it is generally treated as a more qualitative measure for metal nanoparticles. This work demonstrates a gold nanorod (AuNR) spectral morphology analysis (SMA) tool, AuNR-SMA, which is a fast and accurate method to extract quantitative information about an AuNR sample’s structural parameters from its absorption spectra. We apply AuNR-SMA in three distinct applications. First, we demonstrate its utility as an automated analysis tool in a high throughput AuNR synthesis procedure by generating quantitative size information from optical spectra. Second, we use the predictions generated by this model to train a machine learning model capable of predicting the resulting AuNR size distributions from the reaction conditions used to synthesize them. Third, we turn this model to spectra extracted from the literature where no size distributions are reported to impute unreported quantitative information of AuNR synthesis. This approach can potentially be extended to any other nanocrystal system where the absorption spectra are size dependent and accurate numerical simulation of the absorption spectra is possible. In addition, this pipeline could be integrated into automated synthesis apparatuses to provide interpretable data from simple measurements and help explore the synthesis science of nanoparticles in a rational manner or facilitate closed-loop workflows.   

\end{abstract}

\section{Introduction}
Gold nanorods (AuNRs) have drawn significant research efforts due to their applications in cancer cell imaging and treatment, particularly in the development of photothermal therapies, \cite{1_Tong2009, 2_Prashant2008, 3_Kang2017, 4_Mallick2013, 5_Choi2012, 6_Mbalaha2019, 7_Sebastián2012, 8_Zhu2010, 9_Amendola2017} Surface Enhanced Raman Spectroscopy (SERS),\cite{9_Amendola2017, 10_Ou2018, 11_Barros2020, 12_Alvarez-Puebla2011} and photovoltaic devices .\cite{9_Amendola2017, 13_Karg2015, 14_Mubeen2014} These applications rely on the localized surface plasmon resonance (LSPR) of AuNRs, which is highly dependent on the shapes and sizes of these nanoparticles,\cite{9_Amendola2017, 26_Link1999, 27_Eustis2005, 28_Ringe2012} . The LSPR of AuNRs is tuned primarily by changing the ratio of length to width, or aspect ratio (AR), of the nanorods\cite{26_Link1999, 27_Eustis2005, 28_Ringe2012}. Applications of AuNRs thus require the synthesis of these particles with high shape yield and narrow size distributions. However, it remains a challenge to quantitatively understand the impact of AuNR synthesis parameters on the resulting AuNRs.\cite{22_Burrows2017} Quantitatively determining the size and shape distributions of AuNRs requires direct measurement using electron microscopy, which is a time, resource and knowledge intensive task,\cite{15_Lee2020, 17_Wang2021} limiting the amount of data that can be gathered to train models to predict synthesis outcomes.\cite{16_Dahl2020} Although progress has been made automating the size measurement of metal nanoparticles from transmission electron microscopy (TEM) images,\cite{17_Wang2021, 67_Lee2020_Statistical, 68_Laramy2015_HighTroughput} the hardware required for the automation of TEM image collection is not widely available.\cite{52_Mulligan2015} Therefore, we have built an automated analysis model which can extract population level size information from absorption spectroscopy as a tool for more scalable analysis of AuNRs. 

 Recent work has found success predicting the sizes of individual AuNRs from single particle absorption spectra\cite{18_Shiratori2021} and has found limited success at matching individual simulated single particle absorption spectra to samples with very narrow size distributions.\cite{19_Soldatov2021} However, it remains a challenge to provide accurate population level size information for experimental spectra from a sample with a wide size distribution of AuNRs or where spherical AuNPs (AuNS) are also present. An analytical method that can determine AuNR size and shape distributions from absorption spectra and overcome issues related to impure samples \cite{29_Slaughter2010, 30_Encina2007} creates new opportunities for developing interpretable models of AuNR synthesis. 
 
 \par
 Here we overcome these challenges by focusing on the regions of the spectrum where degeneracies from AuNS and left over unreacted growth solution are minimized and selecting spectra according to a series of internal uncertainty metrics. We also enable the prediction of population-level statistics by fitting a 2D gaussian ensemble of simulated AuNR spectra directly to the measured spectrum. This is done using an optimization model which minimizes the difference between an ensemble simulated spectrum and the experimental spectrum. This approach provides much more information on the AuNR sample than existing techniques, which focus on matching single particle simulated spectra to an experimental spectrum. 
\par
We showcase the utility of our automated spectral AuNR morphology analysis in three applications:
\begin{enumerate}
\item automating the analysis of one-pot seedless high throughput AuNR synthesis,
\item training machine learning models to predict AuNR synthesis outcomes,
\item imputing quantitative synthesis data from literature spectra for which population level size data was not reported.
\end{enumerate}
The first two applications comprise a highly active field of research, and AuNP synthesis has already been developed to include prediction of absorption spectrum from synthesis condition, and to rationally design conditions for a specified spectrum.\cite{23_Salley2020, 55_Jiang2022, 56_Vaddi2022, 53_Yazdani2021} However, these procedures focus on relating the conditions to the resulting spectrum\cite{23_Salley2020, 55_Jiang2022, 56_Vaddi2022} or on AuNS synthesis and analysis.\cite{53_Yazdani2021} This limits the abilities of these procedures to produce fundamental synthesis knowledge about how synthesis conditions lead to changes in AuNP morphology, particularly under new reaction conditions. More broadly, there is very limited quantitative information regarding how AuNR synthesis conditions change the resulting AuNR sizes. In recent years, a few studies\cite{22_Burrows2017,23_Salley2020} have attempted to fill in this gap, but these have focused on the seed-mediated synthesis method for AuNRs. Therefore, we focus on the one pot, or seedless\cite{24_Jana2005}, synthesis procedure, and combine the AuNR spectral analysis method we developed with liquid handling synthesis and characterization robots to build an automated synthesis and analysis pipeline. We then utilize this pipeline to build quantitative understanding of the seedless AuNR synthesis procedure and train machine learning models capable of predicting sizes and aspect ratios from initial concentrations of reagents. Using this model, we demonstrate the rich quantitative information that can be extracted from simple measurements and how this information can be applied to nanomaterials discovery. 

\begin{figure}[htbp]
    \centering
    \includegraphics[width=\textwidth]{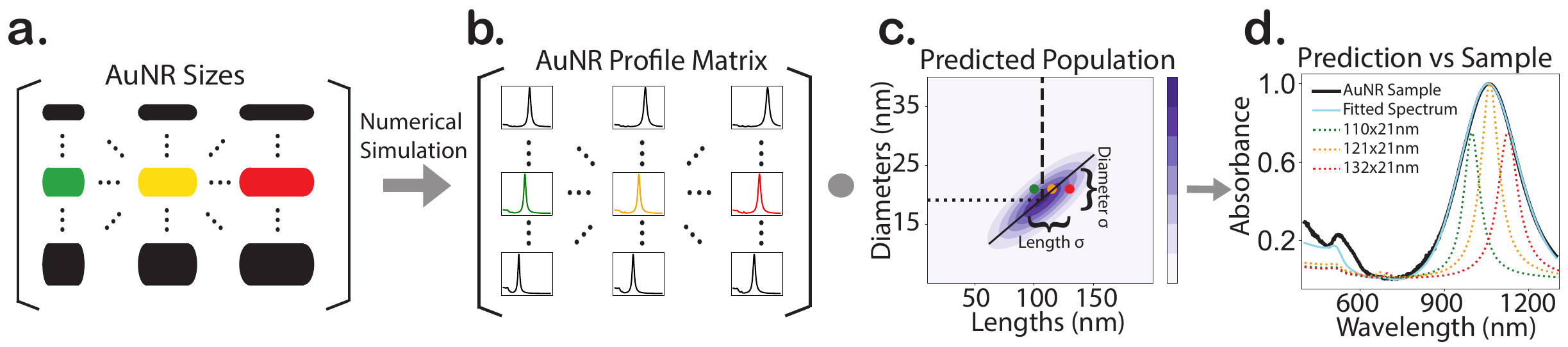}
    \caption{An outline of the spectral morphology analysis, showing how single particle spectra are simulated for a wide range of AuNR sizes, and how these are combined using an inputted distribution to build a simulated mixture spectrum. More details on this process can be found in the SI, section "Spectral Morphology Analysis Outline".}
    \label{model_outline} 
\end{figure}

\section{Results and Discussion}
\subsection{Automated AuNR Size Distribution Analysis} 
UV/Vis/NIR absorption spectroscopy was selected as the analytical method for size distribution inference due to the ease of high throughput characterization and the ability of absorption spectroscopy to simultaneously measure the absorption of every particle in the colloid. Additionally, absorption spectroscopy can be used without sample purification, which can significantly alter the sample's size distribution. Our spectral morphology analysis model uses numerically simulated absorption spectra of single AuNRs to determine the length, diameter and aspect ratio distributions of simulated AuNRs that most accurately reproduce the experimental spectrum. As such, this model can provide not only the mean lengths and aspect ratios of the sample, but also ensemble level information. A full schematic of this model is shown in Figure \ref{model_outline}. 

\par
\begin{figure}[htbp]
    \centering
    \includegraphics[width=\textwidth]{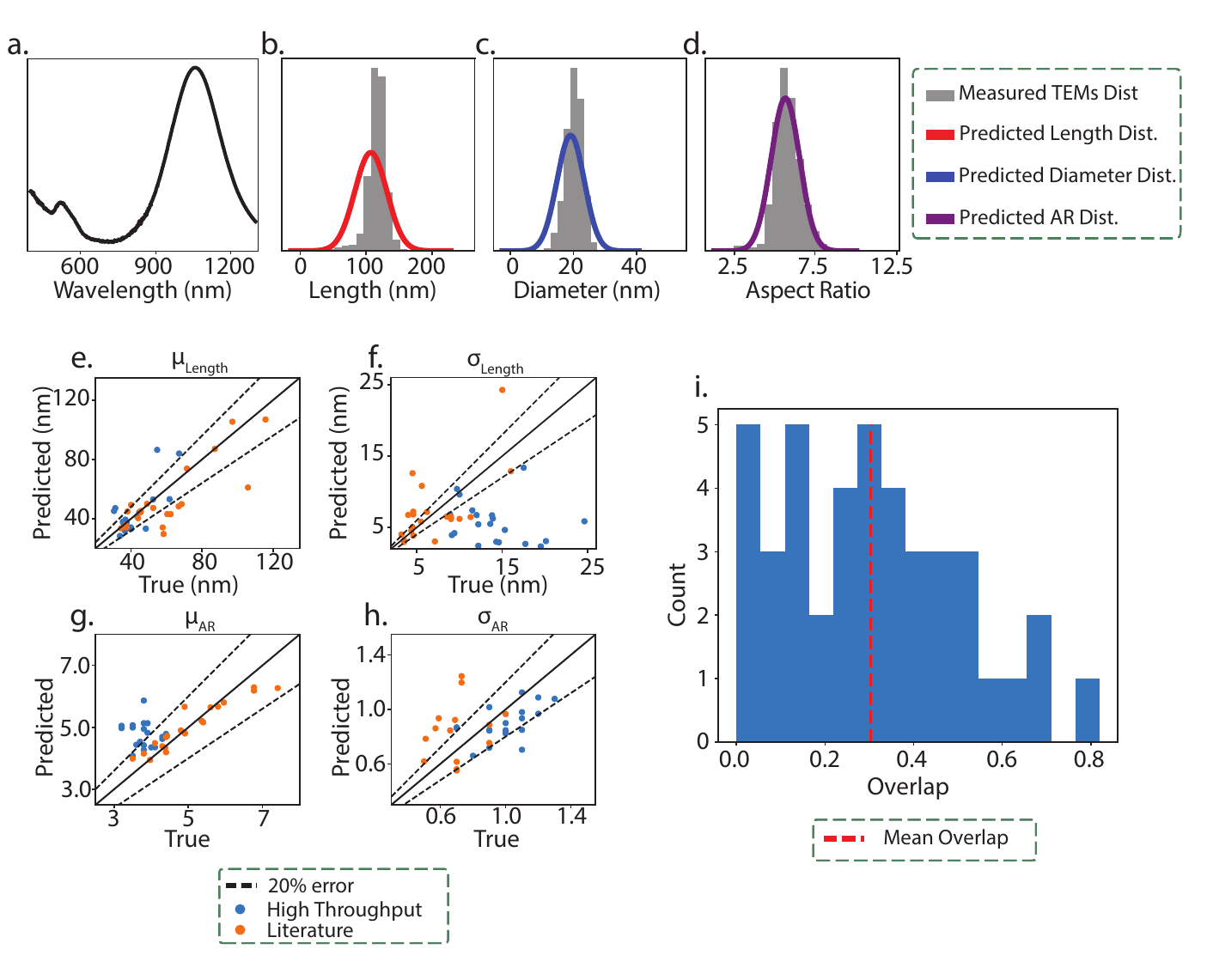}
    \caption{Accuracy of the spectral morphology analysis. (a-d) shows a sample AuNR spectrum (a) and an example of the model output in 1D (b-d). The predicted length distribution (b) is shown in red, the predicted diameter distribution (c) is shown in blue and the predicted aspect ratio distribution (d) is shown in purple. The predicted distributions are plotted over histograms produced by manually measuring AuNRs taken from TEM images. For our 41 validation samples, the accuracy of their size parameters is shown in (e-h). The predictions are colored by whether they came from our high throughput samples (blue) or literature spectra (orange).  $\mu_{Length}$ (e) and $\sigma_{Length}$ (f) are the length mean and standard deviation, respectively.  $\mu_{AR}$ (g) and $\sigma_{AR}$ (h) are the aspect ratio mean and standard deviation, respectively. (i) shows a histogram of the overlap between the predicted 2D distribution in length and diameter space and the true distribution, with the mean overlap indicated with a dashed red line.}
    \label{model_accuracy} 
\end{figure}

 The predictions of this model were validated by TEM analysis of 20 samples from a high throughput synthesis experiment and 21 spectra extracted from literature. Using AutoDetect-mNP \cite{17_Wang2021}, we determined the size and shape distributions of AuNRs in each sample and compared them to the predictions generated from our model. The predictions of the literature spectra were compared to the size distributions reported in the paper. The accuracy of the predictions on our 41 validation spectra are shown in Figure \ref{model_accuracy}.

\par Figure \ref{model_accuracy} shows an example set of outputs provided by this model and their validation against known measurements in 1D. Figure \ref{model_accuracy}a-d shows a sample AuNR spectrum (a) and the predicted length (b) diameter (c) and aspect ratio (d) distributions plotted over histograms produced by manually measuring AuNRs taken from TEM images. The model outputs each of these 1D predictions, as well as the 2D prediction in length and diameter space shown in Figure \ref{model_outline}c. Using the size parameters produced by the model, mean and standard deviation of lengths, diameters and aspect ratios, we can determine the accuracy of this model by comparing to measured TEM images taken in this work or presented in the literature. Figure \ref{model_accuracy}e-h shows these comparisons for length mean (e), length standard deviation (f), aspect ratio mean (g) and aspect ratio standard deviation (h).  

\par As the distributions shown in Figure \ref{model_accuracy}b-d are challenging to use as a model accuracy metric over an ensemble validation set, in this work we express the accuracy as a comparison between the predicted 2D distribution of lengths and diameters and the measured distribution. We do this by calculating an overlap coefficient\cite{65_Madhuri1994_overlap} between the predicted size distribution and the true distribution in length and diameter space (Figure S9). In the case of our high throughput samples, where manual TEM measurements are available, the true distribution was projected onto a normal distribution to ensure accurate comparison with literature distributions, where only means and standard deviations are reported. For literature spectra, where the individually measured TEM particles are not available, the overlap was determined by calculating a distribution based on the reported AuNR means and standard deviations for length and diameter. Using this validation, this model has been shown to produce a predicted distribution which has a high degree of overlap. The average overlap across all the validation spectra is shown to be around 0.3 in Figure \ref{model_accuracy}, showing that, on average, this model is able to reasonably reproduce the population distribution of AuNRs present in the colloid (Figure S7, S9).  
 
\subsection{Application 1 - High Throughput Synthesis Size Mapping} 
\begin{figure}[htbp]
    \centering
    \includegraphics[width=\textwidth]{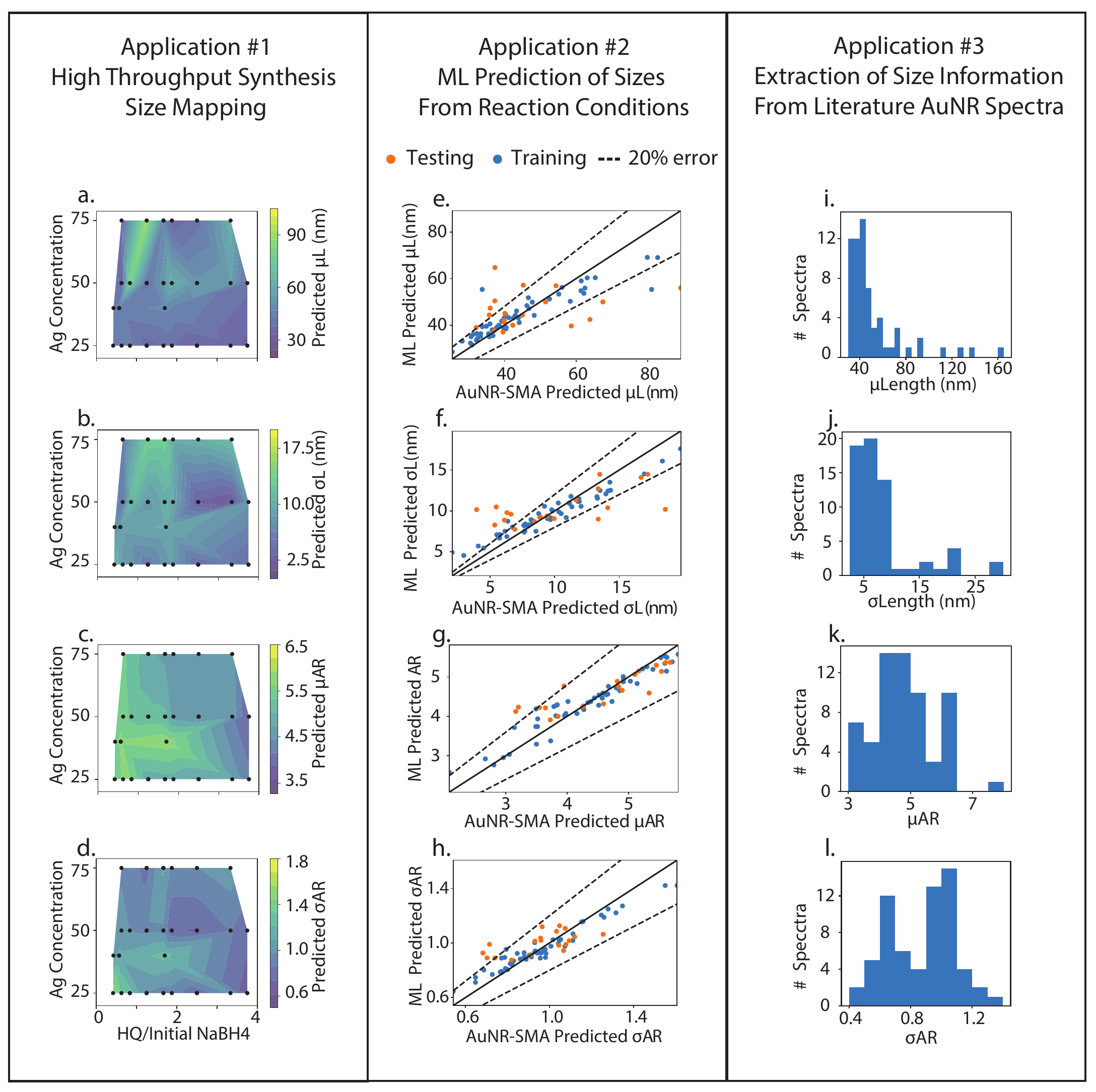}
    \caption{Application 1 - 2D representations of the impact of changing synthesis conditions on the produced length mean (a), length standard deviation (b) aspect ratio mean (c) and aspect ratio standard deviation (d) generated using our model. The contour plots show how each of the four size parameters change with the ratio of hydroquinone to initial NaBH\textsubscript{4} (x axis) and the AgNO\textsubscript{3} concentration (y axis). The black dots show where experiments have been conducted that produced spectra of sufficient quality to yield a prediction. Application 2 - Accuracy of the ML model trained on the spectra labeled with sizes generated by our high-throughput experiments and size analysis model. (e) length mean (f) length standard deviation (g) mean aspect ratio (h) aspect ratio standard deviation. Training accuracy is shown in blue while accuracy on test data is shown in orange. Application 3 - Distributions of size parameters from the 64 unlabeled literature spectra predicted in this work. (i) length means (j) length standard deviations (k) mean aspect ratios (l) aspect ratio standard deviations.}
    \label{model_applications} 
\end{figure}

By systematically varying concentrations of AgNO\textsubscript{3}, hydroquinone, initial NaBH\textsubscript{4} and total NaBH\textsubscript{4} (See Methods Section - High-Throughput Synthesis), and utilizing our automated spectral morphology analysis on the spectra of the resulting samples, we were able to determine trends in sizes and size distributions of the AuNRs. This high throughput synthesis reaction produced 48 samples where rods were successfully synthesized. Of these 48 samples, 6 spectra could not be processed by our spectral morphology analysis due to high prediction uncertainty or a longitudinal peak at too high energy (Figure \ref{model_flowchart}). Using the successfully fit samples, a synthesis map of the size parameters predicted from their absorption spectra is shown in Figure \ref{model_applications}a-d. The x axes shows the ratio of hydroquinone to the first addition of NaBH\textsubscript{4}, which was chosen due to the combination of these two conditions controlling the growth kinetics, which has been shown to be responsible for changing morphology in AuNR synthesis\cite{57_Liu2017}. Additionally, interactions with the weak and strong reducing agents have also been shown to impact the sizes and yield of AuNRs\cite{22_Burrows2017}. AgNO\textsubscript{3} concentration was visualized on the y axes due to myriad observations that AgNO\textsubscript{3} concentration is essential to controlling AuNR morphology \cite{22_Burrows2017, 61_Nikoobakht2003, 62_Jackson2014}. 

\par The predicted lengths show that at low AgNO\textsubscript{3} concentration the maximum length produced with any combination of hydroquinone and NaBH\textsubscript{4} produces AuNR lengths from 30-50nm. However, at higher concentrations of AgNO\textsubscript{3}, lengths of 70-90nm can be produced for specific hydroquinone and NaBH\textsubscript{4} ratios. As these resulting sizes are much less common for this synthesis procedure, one may be inclined to attribute this less common result to an error in the spectral morphology analysis. However, one of the two sets of conditions resulting in longer AuNRs contains two samples with the same ratio of reaction conditions, and those produced mean length predictions that were less than 10 nm apart, making it unlikely that a simple large error in the model is responsible for this result. Overall, we demonstrate that our model can be integrated into high throughput experiments as an automated size and size distribution prediction tool, generating detailed quantitative information on the fly. This allows for rapid detailed analysis and enables the possibility of closed loop workflows for robotic AuNR synthesis. 

\subsection{Application 2 - ML Prediction of Sizes From Reaction Conditions} 
Besides enabling more quantitative understanding of AuNR synthesis, the ability to predict synthesis outcomes is an important goal to enhance synthetic exploration and optimization. By combining our synthesis and analysis pipeline with machine learning, we predict the averages and standard deviations of lengths and aspect ratios of AuNRs synthesized with our one-pot procedure from their synthetic parameters. We use a dataset containing 73 samples, where their reaction conditions were labeled by their resulting AuNR sizes predicted from our model. We use the following algorithms, as implemented in scikit-learn\cite{64_sklearn}, to attempt to predict the output parameters from synthetic inputs: Elastic Net, Bayesian Ridge Regression, Kernel Ridge Regression, Support Vector Regression, Random Forest Regression and Gradient Boosting Regression. Generally, random forest regression models outperformed k-nearest neighbor, support vector regression, kernel ridge regression and bayesian ridge regression and elastic nets for most of this dataset. Gradient boosting regression delivered similar results, though support vector regression performed better for predicting lengths and k-nearest neighbor models performed best for aspect ratio standard deviations (See Table S2). The models were reasonably well predictive for aspect ratio and standard deviations of aspect ratios, with test R$^2$ values of around 0.7 and $\tilde{} $10\% errors (Figure \ref{model_applications}e-h). While length and length standard deviations were less well predicted, error ranges were  similar to those of the spectral morphology analysis. Previous work using machine learning has shown the ability to predict spectral alignment of AuNRs in seeded growth synthesis \cite{23_Salley2020, 55_Jiang2022}. We show here that it is possible to integrate the information gained from spectral morphology analysis with machine learning methods to directly predict the size and aspect ratios of AuNRs from their synthesis conditions in a one-pot, seedless growth procedure. In principle, this analysis routine could also be used in tandem with kinetic models to produce scientific machine learning models of particle growth.\cite{63_DAHL} 

\subsection{Application 3 - Extraction of Size Information From Literature AuNR Spectra} 
In addition to this model's utility as an automated analysis tool for AuNR synthesis and our ability to generate predictions of resulting AuNR morphology from reaction conditions, we also turn this model to the AuNR synthesis literature to impute size information from published spectra without direct size measurements. Our extraction produced 64 unlabeled literature spectra that were successfully fit by our model (Figure \ref{model_applications}i-l). The most common aspect ratios reported in the literature were observed between 4 and 5, with a non negligible spread higher into ARs of 5-6, although very few are seen greater than 7. Similarly, AuNR lengths of 30-70 nm are far more commonly seen in the literature than longer lengths, although rods greater than 100 nm have been successfully synthesized. Looking in detail at three synthetically interesting outliers that are predicted to have lengths > 100 nm and where we found TEM images of similar synthesis outcomes in the same report\cite{25_Vigderman2013,40_Zhang2014,69_Wang2016}, the first comes from a publication reporting syntheses that result in rods with length > 100 nm \cite{25_Vigderman2013}, the second most likely has a length of roughly 90 nm\cite{69_Wang2016} and the third has TEM images showing rods with roughly 50 nm length \cite{40_Zhang2014}. A further detailed comparison and discussion of potential external factors that can induce errors in predictions can be found in the SI. Overall, we demonstrate that our model can be used to impute missing size and size distribution information from previous literature reports, contributing to a richer data ecosystem, allowing more in-depth analysis, and enabling machine learning based on previous literature results.

\section{Conclusion}
In this work, we have developed a physics based automated spectral morphology analysis tool for colloidal AuNR samples. This model is able to predict the AuNR’s length and aspect ratio mean and standard deviation with a high degree of accuracy, producing population level information on the produced AuNR sample from its absorption spectrum. We validated this model by comparing its predictions to measured AuNR size distributions extracted  from transmission electron microscopy images of the predicted AuNR samples and spectra labeled with size distributions extracted from the literature. We show the wide range of applications this model can be used for by utilizing it as an automated analysis tool in high throughput AuNR synthesis, to generate data for training machine learning models to predict synthesis results based on reaction conditions, and as a literature mining tool to increase the volume of quantitative knowledge available on AuNR synthesis. Through these applications, we elucidate quantitative information about seedless AuNR synthesis using hydroquinone and the relationships between the reaction conditions and the size outcomes. In principle, this approach can be extended to any other nanocrystal system where the absorption spectra are size dependent and accurate numerical simulation of the absorption spectra is possible.

\section{Methods}
\subsection{Materials} 
Hexadecyltrimethylammonium bromide (CTAB, 98.0\%) was purchased from TCI America. Hydrogen tetrachloroaurate trihydrate (HAuCl\textsubscript{4}·3H\textsubscript{2}O, 99.9\%), silver nitrate (AgNO\textsubscript{3}, 99.0\%), and sodium borohydride (NaBH\textsubscript{4}, 99.99\%) were obtained from Sigma-Aldrich. Diglycol methyl ether (diglyme 99.5\%) was used to dissolve NaBH\textsubscript{4} and was purchased from  Sigma-Aldrich. Hydroquinone (99\%) was purchased from Sigma-Aldrich. NaBH\textsubscript{4} powder was stored in a nitrogen  glovebox. HAuCl\textsubscript{4}·3H\textsubscript{2}O, hydroquinone, and AgNO\textsubscript{3} were stored in a vacuum desiccator at room temperature. Deionized water was used for all aqueous solutions. All chemicals were used without further purification.

\begin{figure}[htbp]
    \centering
    \includegraphics[width=\textwidth]{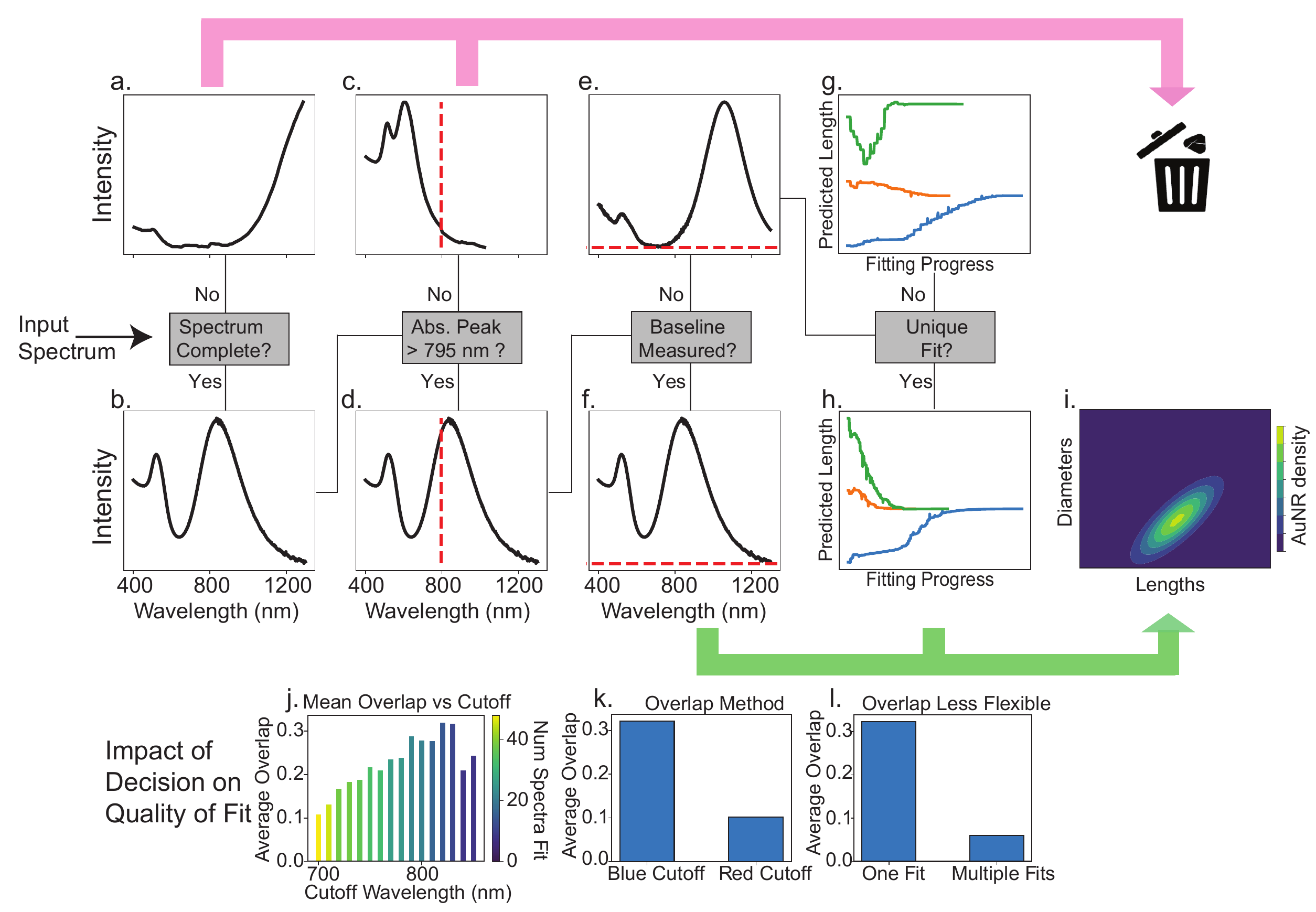}
    \caption{The steps the overall spectral morphology analysis uses to determine how, or if, an input spectrum is fit and whether the results of that fit are expected to be accurate. Briefly, the model first determines whether the inputted spectrum is complete (a-b) discarding incomplete spectra. The model then determines if the longitudinal peak is before or after 795 nm (c-d) discarding spectra with peak wavelengths shorter than 795 nm. Finally, the model determines whether the spectra has been measured out to the baseline (e-f) and uses a different fitting procedure in each of these cases (g-h). The model then outputs a 2D distribution of the AuNRs fitted to the sample (i). The rational for selecting the 795 nm cutoff is shown in (j) and the validity of the differing methods for the measured baseline and the discarding of uncertain spectra are shown in (k-l). The overlap metric used in (j-l) is the overlap of the predicted AuNR population and the measured AuNR population using TEM, and a full description of this metric can be found in Figure S9. A detailed description of the full model workflow can be found in the SI, section "Spectral Morphology Analysis Details".}
    \label{model_flowchart} 
\end{figure}

\subsection{Automated Size Prediction from Absorption Spectra} 
AuNR morphology prediction from the absorption spectrum of a colloidal sample is based on basic optical properties of colloidal samples of AuNRs. First, the overall spectrum can be approximated as a sum of the absorption of each particle in the colloid, provided nothing else is absorbing and/or scattering in the energy range of the spectrum. It then becomes possible to determine the sizes of AuNRs present in a sample by fitting the relative contribution of AuNRs of different sizes and finding the distribution which best rebuilds the sample’s spectrum. These assumptions serve as the basis of our spectral morphology analysis. 

The basis set for spectral fitting in our model consists of numerically simulated single particle spectra of AuNRs from the scuff-em simulation procedure, which simulates absorption spectra using boundary-element methods, as shown in Figure 
\ref{model_outline}a and b.\cite{21_Elzouka2020} The accuracy of these simulations, and their ability to accurately reproduce a full experimental absorption spectrum, has been shown by comparison with TEM measurements (Figure S3). The spectral morphology analysis uses these single particle spectra to reproduce a colloid’s absorption spectrum by generating a size distribution based on AuNR mean length/standard deviation, mean diameter/standard deviation and the correlation between the length and diameter distributions. This distribution is then applied to the basis set, building a spectrum of a sample with that underlying size distribution, as shown in Figure \ref{model_outline}c and d. This procedure then returns the values of the 5 parameters that build the best fit, indicating the predicted population of AuNRs in this sample, as well as the chi square between the fitted and experimental spectrum as a measure of how well the fit has replicated the experimental spectrum. In cases where there are multiple stable fits, unless specified otherwise, the prediction is the fit that has the lowest chi square value between the two spectra. The accuracy of this model at predicting size distributions across a broad range of size parameters has been validated by manual TEM analysis by our group and extracted spectra from literature results on AuNR synthesis.\cite{31_Requejo2017,32_Zhang2016,33_Scarabelli2013,34_Caño2020,35_Requejo2018,36_Requejo2020,37_Wei2021,38_Lai2014,39_Ye2013,40_Zhang2014,41_Gallagher2021,42_Link1999,43_Xu2014,44_Salavatov2018,45_Ye2012,46_Ye2013,47_Roach2018,48_Huang2018,49_Tong2017,50_Liopo2015,51_Li2019} 

\subsection{Automated Size Prediction Model Validation}
The accuracy of our size prediction model was validated using samples generated from a high thorughput AuNR synthesis run and spectra extracted from the literature. Our high throughput synthesis procedure produced 20 samples deemed suitable for validation. This required at least moderate shape purity, as many reaction conditions in our high throughput experiment produced a significant number of AuNS. This was determined by setting a threshold that the ratio between the transverse plasmon peak at roughly 520 nm, where AuNS absorption occurs, and the longitudinal peak intensity had to be less than one. Additionally, only samples following the filtering criteria set out in Figure \ref{model_flowchart} were fit. All of the high throughput samples fitting the selection criteria were fit successfully. At least 500 particles were measured for each high throughput synthesis sample by taking between 50 and 100 TEM images. The particles were then measured using AutoDetect-mNP, an automated size and shape detection algorithm for TEM images of metal nanoparticles.\cite{17_Wang2021}

In the case of literature spectra, 86 spectra were extracted with reported size information. Of these, 19 were discarded due to either highly irregular synthesis conditions deemed likely to impact the spectra (ie large volumes of biological additives) or noted systematic errors in spectral processing (spectra that didn't baseline subtract the H2O peak in the NIR, etc). 12 were discarded due to the spectra being incomplete, determined by the longitudinal peak ending before the FWHM of the peak. 10 were discarded due to the reported size parameters being outside the range of the lower bound of our diameters, denoted by a threshold set 1 nm higher than our lowest diameter value, 5 nm. This threshold was chosen due to observations that fitted populations with the mean diameter lower than 6 nm often had appreciable density in regions where there were no simulated spectra. 4 additional spectra were discarded due to their longitudinal peak being too high in energy(see Figure \ref{model_flowchart})c-d. 17 were returned as too uncertain to be fit (see Figure \ref{model_flowchart})g and an additional 3 failed due to the values hitting the bounds of the prediction space. This resulted in 21 literature spectra that were added to our validation set of labeled spectra. The full validation set contained 41 labeled spectra, which comprise Figure \ref{model_accuracy}. 

\subsection{Manual AuNR Synthesis} 
AuNRs were prepared using a modified version of the procedure described by Vigderman et al.\cite{25_Vigderman2013} In a general synthesis, aqueous solutions of CTAB, HAuCl\textsubscript{4}, and AgNO\textsubscript{3}, in that order, were added to a 1 mL glass vial with a stir bar. Then, an aqueous solution of hydroquinone was added to the reaction mixture to reduce Au(III) to Au(I), inducing a color change from yellow to clear. After waiting 10 minutes, a solution of NaBH\textsubscript{4} dissolved in diglyme and sonicated for 10 minutes was added to the solution under rapid stirring to induce the formation of AuNRs. Stirring was then stopped after half an hour, and the solutions were allowed to grow overnight to ensure the reaction had completed. Then, each solution was centrifuged at 10000 rpm for 30 minutes and re-suspended in DI water under gentle sonication to disperse the pellet.  

\subsection{Absorbance Spectroscopy} 
UV–vis absorption spectroscopy of the synthesized AuNRs was conducted using a Shimadzu UV-3600 double beam spectrometer. Samples were prepared by re-suspending the centrifuged sample in 1 mL of DI water and then diluting 300 uL of this sample with 2.7 mL DI water. Before measurements, a background spectrum was recorded and subtracted using a cuvette filled with 3 mL of DI water. 

\subsection{Transmission electron microscopy (TEM)}
Images of the AuNR samples were taken with a FEI Tecnai T20 transmission electron microscope equipped with a Gatan RIO16IS camera and a LaB$_6$ filament. All images were recorded under 200 kV accelerating voltage. Samples were prepared for TEM by drop casting from the 3 mL sample used for absorption spectral analysis, which was found to have the appropriate concentration, onto a carbon support with 400 copper mesh. Samples were dried overnight by placing them under vacuum. 

\subsection{High-Throughput Synthesis}
High throughput synthesis of AuNRs was performed using a Hamilton Microlab NIMBUS4 liquid handling robot. For high-throughput synthesis, new aqueous solutions of CTAB (100 mM), Hydroquinone (60 mM), HAuCl\textsubscript{4} (4 mM) AgNO\textsubscript{3} (5 mM) and NaBH\textsubscript{4} (0.3 mM) were prepared for each run. These solutions were then added to single use 1 mL glass vials in the order DI H\textsubscript{2}O, CTAB, HAuCl\textsubscript{4}, AgNO\textsubscript{3}, Hydroquinone, NaBH\textsubscript{4} while the entire plate was heated to 30 \textdegree C and shaken at 300 rpm. Shaking continued for half an hour after the final addition was completed, and the plate was then left undisturbed for 4 hours. The samples were then removed from the plate and placed in a dark cabinet and wrapped in aluminum foil to continue reacting overnight. 

Several components of the typical hydroquinone seedless synthesis procedure\cite{25_Vigderman2013} were changed to produce a reaction that was compatible with our liquid handling synthesis robot setup. The overall scale of the reaction was significantly decreased to ensure a 96 well plate could be used for high-throughput synthesis, optimizing the number of reactions achieved with each run. This necessitated the use of 1 mL vials, producing 0.5 mL of sample, which is roughly a factor of 20 below the volume produced by the previously reported procedures. Additionally, to maximize the liquid handling precision of the NIMBUS robot, the minimum volumes for each reagent are maintained at 5 $\mu$L, necessitating a lower concentration of stock solutions such as NaBH\textsubscript{4}. Diglyme was chosen as a polar, aprotic solvent to eliminate background reactions of NaBH\textsubscript{4} with H\textsubscript{2}O. This work found NaBH\textsubscript{4} to be stable in diglyme for over 24 hours (Figure S8). The enhanced stability also allows for more flexibility with the reaction procedure, allowing reactions to be performed with multiple addition of NaBH\textsubscript{4} and potentially allowing options for more advanced procedures such as AuNR synthesis with a syringe pump. Finally, the stability of the stock solution allows for improved reproducibility of the synthesis procedure, potentially overcoming a chronic problem in the field (Figures S4, S8).

\begin{acknowledgement}

Work at the Molecular Foundry was supported by the Office of Science, Office of Basic Energy Sciences, of the U.S. Department of Energy under Contract No. DE-AC02-05CH11231. Other work reported here was supported by the U.S. Department of Energy, Office of Science, Office of Basic Energy Sciences, Materials Sciences and Engineering Division, under Contract No. DE-AC02-05-CH11231 within the Data Science for Data-Driven Synthesis Science grant (KCD2S2). Computational resources for simulations of AuNR spectra were provided by the Lawrencium computational cluster resource provided by the IT Division at the Lawrence Berkeley National Laboratory with funding by the Director, Office of Science, Office of Basic Energy Sciences, of the U.S. Department of Energy under Contract No. DE-AC02-05CH11231. J.C.D acknowledges funding through the NSF-GRFP program under DGE 1752814 and the Kavli NanoScience Institute, University of California, Berkeley through the Philomathia Graduate Student Fellowship.

\end{acknowledgement}

\section*{Data and Code Availability}
The dataset of simulated AuNR spectra, the high throughput synthesis reaction conditions, resulting spectra, size predictions and TEM images used for analysis, literature spectra, size predictions, true sizes and reaction conditions (when applicable) and the code generating the size prediction model will be shared upon request.   

\section*{Author Contributions} 
SPG developed, validated and applied the automated size prediction model, developed the high throughput synthesis procedure,  conducted the high throughput experiments and wrote the manuscript. JCD provided training, expertise, and conducted high throughput synthesis experiments, advising on the automated size prediction model construction and developed the machine learning models discussed in Application 2. ME generated the simulated AuNR spectra. XW provided automated size prediction modeling for the TEM images taken in this work. DOB took the validation TEM images necessary to validate the size prediction model. MG and HC extracted and labeled literature spectra. RP and SL provided advising and expertise related to the simulated AuNR spectra generation. EC provided advising, expertise and training for the high throughput AuNR synthesis experiments. APA provided experimental AuNR synthesis knowledge, project planning, led the collaboration and designed the scope of this work. All authors read, edited and approved the final manuscript.

\bibliography{main}

\newpage

\section{Supporting Information}

\subsection{Spectral Morphology Analysis Outline}
Our basis set for the spectral morphology analysis contains particles set on a 1nm scale for ease of interpretability (Figure \ref{model_outline}a). Using the Scuff-em simulation package \cite{21_Elzouka2020}, the single particle absorption spectra of each of these AuNRs is simulated and stored in a matrix of identical dimensions to the one with the size information labels (Figure \ref{model_outline}b). When an individual spectrum is predicted, a population distribution is generated based on predicted size parameters (Figure \ref{model_outline}c). The fitted size parameters are the length mean and standard deviation, the diameter mean and standard deviation, and the correlation between the two. This generates a two dimensional probability distribution which is set on the same scale as the previous two matrices and normalized. Each entry in this distribution can then be treated as the weight that each individual single particle spectrum contributes to the simulated mixture spectrum. By multiplying each weight by its corresponding single particle spectrum and summing the resulting spectra, a simulated mixture spectrum is developed and compared to an unknown sample (Figure \ref{model_outline}d). This process is then repeated until the size parameters have been found which generate a mixture spectrum that most closely matches the unknown sample spectrum. This iterative process changes the parameters which build the simulated mixture spectrum, AuNR length mean and standard deviation, AuNR diameter mean and standard deviation, and the correlation between these two, until it finds a set of values for these 5 parameters that minimizes the differences between the simulated mixture spectrum and the sample spectrum. To ensure that the returned sets of size parameters reflect the global minimum and not simply a local minimum, mean lengths of 50, 100, and 150 nm and standard deviations from 10 to 20 percent are seeded as initial guesses. The diameter is set by setting the initial aspect ratio as predicted by the equation from\cite{25_Vigderman2013} that gives the aspect ratio from the spectrum as LSPR Peak = 420 + 95AR. Although plasmon coupling could undermine the accuracy of the additive model, its effect is negligible in most synthesized colloidal systems, and is therefore safely assumed to be not present\cite{20_Jain2010}. If the fit hits one of the imposed bounds, which are set to prevent illogical values being predicted (i.e. distributions that would result in negative sizes), the fit is labeled as a failed fit.

\subsection{Spectra Processing and Smoothing} 
Spectra are smoothed using a savitzky-golay filter implemented using scipy's signal module. The smoothing parameters are based on whether the spectrum came from the NIRVANA instrument, from our individual shimadzu, or extracted from literature. As the NIRVANA instrument has essentially no noise, the smoothing parameters use a window size of 5 and a polynomial order of 3. literature extraction spectra are also less noisy on average, resulting in a set of smoothing parameters using a window size of 51 and a polynomial order of 3. Our manually taken spectra have higher noise due to the lower concentration of AuNRs in our samples when they are diluted to the volume needed for our instrument, so in this case the window size is increased to 101. All sample spectra are baseline subtracted and normalized to one post smoothing. 

\subsection{Spectral Morphology Analysis Details} 
This procedure filters out spectra where the longitudinal peak is lower in intensity than the transverse peak, which indicates a high volume of spherical impurities, spectra where less than the FWHM of the spectrum is recorded on the red edge and also spectra where the longitudinal peak is redder than 795 nm. This ensures that no absorption from spherical particles or any unreacted growth solution still in the sample convolutes the spectrum. As seen in Figure S5, the presence of these quantities can change the spectrum and confuse the automated spectral morphology analysis, which assumes that all the intensity in the spectrum is originating only from AuNRs of different sizes. Attempts to include spheres and growth solution into the fitting produced a series of inaccurate results, and therefore such procedures were discarded in favor of only utilizing the section of the spectrum that can be confidently assumed to be only AuNR absorption. The threshold of 795 nm was chosen by examining the model’s accuracy on a set of validation spectra where the sizes were known at cutoffs ranging from 700-850 nm and weighing this against the number of spectra the model is able to predict at that cutoff, since further red shifting the cutoff will decrease the number of spectra the model is able to predict. 795 nm was found to have the third highest accuracy and predict nearly 75\% more of our validation spectra than the two higher cutoffs, 825 and 835. This is shown in Figure \ref{model_flowchart}j. After this filtering step, the model classifies spectra into those where the red edge goes to the baseline, as determined by the minimum point being on the red edge of the spectrum, and those where the baseline is the region in between peaks, as determined by the minimum point being between the two peaks. If the spectrum runs to the red edge enough for the baseline to be established on the red edge, and therefore we can assume it is the true baseline of the spectrum, the model is run without further amendments and the prediction which produces the lowest chi square and avoids hitting any bounds is taken as the final fit. If, however, the baseline is found to be between the two peaks, the minimum of the region between the transverse and longitudinal in the simulated spectrum is subtracted off of the spectrum as a whole to make this region the baseline of the simulated spectrum. Additionally, the blue edge of the spectrum is cut off at 650 nm rather than 795, to ensure that the model has one edge of the baseline of the extracted longitudinal peak. This has been observed to be necessary for an accurate prediction. Then the model is fit using a less flexible model, where only 4 parameters are fit independently: Length mean, diameter mean, a relative standard deviation parameter that is the same for length and diameter, and the 2d correlation. This forces length and diameter to have the same relative standard deviation, rather than fitting them both independently. If this model returns a prediction with only one stable fit, as demonstrated in Figure \ref{model_flowchart}h, that is accepted as the prediction. If, however, there are multiple or zero stable fits, as demonstrated in \ref{model_flowchart}g then the fit is run again with a more flexible model, the original 5 parameter length/diameter standard deviation correlation model, and then if this model returns a prediction with only one stable fit, that is accepted as the prediction. If, however, there are multiple or zero stable fits, the spectrum is discarded as a failed fit. At each step of this multilayered procedure, the accuracy, as determined by the average overlap between the predicted size distribution and the true size distribution, was compared to the accuracy of omitting that individual step, as shown by Figure \ref{model_flowchart}k-l, to justify this approach. 

\par The predicted mean aspect ratio, shown in Figure \ref{model_accuracy}g, shows a distinct difference in the level of accuracy for the high throughput synthesis samples and samples extracted from literature. This difference can be explained by the use of hydroquinone as the weak reducing agent in the high throughput procedure, which has been known to make AuNRs with flattened tips, rather than the nearly perfectly hemispherical tips made by the more common procedure using ascorbic acid as the weak reducing agent\cite{25_Vigderman2013}. This flattening of the AuNR’s tip causes a red shift in the longitudinal peak, leading the spectral morphology analysis to overestimate the sample’s aspect ratio, as shown in literature\cite{25_Vigderman2013} and the simulations created here (Figure S1). The ascorbic acid procedure is much more common across the samples extracted from the literature, explaining why the aspect ratio is much more accurately predicted with our literature validation samples. \cite{31_Requejo2017,32_Zhang2016,33_Scarabelli2013,34_Caño2020,35_Requejo2018,36_Requejo2020,37_Wei2021,38_Lai2014,39_Ye2013,40_Zhang2014,41_Gallagher2021,42_Link1999,43_Xu2014,44_Salavatov2018,45_Ye2012,46_Ye2013,47_Roach2018,48_Huang2018,49_Tong2017,50_Liopo2015,51_Li2019} However, despite synthetic differences that change the rod's refractive index, the spectral morphology analysis can still attain a reasonable overlap. The predictions of the mean length, shown in Figure \ref{model_accuracy}e, are more consistent between high throughput and literature samples, and shows the vast majority of the predictions are within 20\% error, a metric beyond the existing state of the art for AuNR size prediction.\cite{18_Shiratori2021, 19_Soldatov2021} The predicted aspect ratio standard deviation, shown in Figure \ref{model_accuracy}h, is quite accurate overall, showing less than 25\% of the validation set samples have an error greater than 20\%. It should also be noted that each of the true values for the aspect ratio and length standard deviations carry inherent uncertainty due to the large number of particles required to build an accurate measured distribution. Qualitative estimates have been made showing that shape identifications in AuNRs can have be incorrect in as much as 10\% of identified particles, implying that absolute size may carry an even larger uncertainty\cite{17_Wang2021} This may explain some of the errors shown in the length standard deviation prediction, shown in the Figure \ref{model_accuracy}f, such as the samples with improbably high standard deviations that are underestimated by the model.  

\subsection{Literature Spectral Extraction and Prediction}

Literature spectra were extracted by utilizing webplotdigitizer's x step with interpolation function \cite{66_Rohatgi2024_webplot}. High resolution images of each plot were downloaded from each paper. 

Figure \ref{model_applications}i shows a few higher length predictions greater than 100 nm. To probe the outliers in this distribution in some detail, the higher length rods predicted by this model were examined - as they are both outliers and synthetically most interesting. Errors in prediction occur most often when unusual reaction conditions are used, which can change the refractive index of the colloid solution or the AuNRs themselves. Of the 4 samples predicted to be over 100 nm in length, one shows TEMs which qualitatively appear to match its prediction quite well. This procedure is capable of producing rods greater than 100 nm in length,\cite{25_Vigderman2013}  and one of the spectra without size information extracted from this paper is predicted to be 113.1 nm. Of the remaining 3, one is inconclusive as only 1 TEM image is included in the paper,\cite{51_Li2019} so its prediction of 161.7 nm length is impossible to confirm. Although this length seems improbably long based on the bulk of other documented and predicted results, this procedure used hydrogen peroxide as a weak reducing agent, which is a relatively unexplored procedure and its morphology bounds are not well understood. One is a mild overestimate, returning prediction of 139.6 nm when the TEMs shown in the paper appear to be roughly 90nm \cite{69_Wang2016}. The final appears to be a larger miss by the model, as TEMs shown in the paper are roughly 50 nm while the prediction is 127 nm \cite{40_Zhang2014}. However, this procedure was conducted at higher temperature, making it possible that the morphology of the resulting AuNRs have a different refractive index than those this model was trained and validated on, potentially explaining this error. The remaining distribution falls within the space of lengths common in AuNR synthesis, making a miss of this magnitude unlikely for the remainder of the samples.

\subsection{Overlap Metric}
Direct comparison of the distributions of length, diameter and aspect ratio, as shown in Figure \ref{model_accuracy}b-d, is a challenging way to visualize accuracy for multiple samples. Therefore the overlap metric was developed to allow the accuracy of the predictions to be distilled into a single quantity. To calculate the overlap between a predicted size distribution and the measured size distribution from TEM analysis, every point in the two distributions, which are both on a 1 nm scale, are compared, and the point that is lower in value is assigned to that point in the overlap matrix (Figure S9 f-g). Upon construction of the overlap matrix, every point is summed together to produce the overlap coefficient (Figure S9). This coefficient will be between zero and one, with a higher overlap indicating a more accurate prediction. 

\subsection{High Throughput Synthesis of AuNRs} 
 The utility of the spectral morphology analysis routine is demonstrated by predicting size distributions of samples generated by high throughput synthesis. In this procedure, NaBH\textsubscript{4} is added directly to the growth solution, rather than first synthesizing gold seeds in a separate step and adding these to the reaction. The NaBH\textsubscript{4} creates seeds in situ, which then serve to grow into AuNRs\cite{58_Samal2010} 
 
 \par Two complete 96 well plates of AuNR synthesis reactions are reported in this work. The first was used to build and calibrate the high throughput synthesis model and the second was used to illustrate its utility. In the first experiment, only the amounts of hydroquinone and CTAB were varied (Figure S2). Figure S1 shows there appears to be an inflection point between 14 and 20 mM CTAB where the samples transition from making virtually no AuNRs to a moderately successful reaction, as shown by the increased intensity of the longitudinal peak relative to the transverse peak between the 14 and 20 mM spectra. Additionally, further increasing the CTAB concentration increases the shape purity and aspect ratios of the samples, although increasing the concentration to 58 mM appears to broaden the aspect ratio distribution significantly (Figure S2), as evidenced by the broadening of the longitudinal peak. For hydroquinone, increasing the concentration to 10.8 mM steadily increases AuNR yield and slightly blue shifts the longitudinal peak, but an increase past this concentration appears to begin to decrease yield, as the ratio of intensities of the longitudinal and transverse peaks decreases past this concentration. The second set of high throughput synthesis results were analyzed by spectral morphology analysis to produce detailed quantitative information regarding the impact of the synthetic condition on AuNR structural properties (see Figure \ref{model_applications}a-d).   

 \subsection{Machine Learning}

Chemical input data and predicted morphology parameters from the spectra for all the high-throughput synthesis were merged, and any unsuccessful analyses were dropped, resulting in a small dataset of 73 samples. Chemical inputs were designated as the amounts of water, gold chloride, silver nitrate, cetyltrimethylammonium bromide, hydroquinone and sodium borohydride added to the reaction solution. For each of the output parameters ($\mu_{Length}$, $\sigma_{Length}$, $\mu_{Diameter}$, $\sigma_{Diameter}$, Aspect Ratio ($\mu_{AR}$), $\sigma_{AR}$), the following algorithms as implemented in scikit-learn\cite{64_sklearn} were used to attempt to predict the output parameters from synthetic inputs: Elastic Net, Bayesian Ridge Regression, Kernel Ridge Regression, Support Vector Regression, Random Forest Regression and Gradient Boosting Regression. For each algorithm, a grid search cross validation was performed using the parameters and values specified in Table S1 below, all other parameters remained at preset default values.

\begin{table}[]
\renewcommand{\thetable}{S\arabic{table}}
\resizebox{\textwidth}{!}{%
\begin{tabular}{llllllllll}
                  & Parameter 1     & Values                           & Parameter 2          & Values                  & Parameter 3        & Values                             & Parameter 4   & Values                    &                   \\
Elastic Net       & 'alpha'         & \multicolumn{3}{l}{{[}0.0001, 0.001, 0.01, 0.1, 1.0, 10.0{]}}                     & l1\_ratio'         & \multicolumn{3}{l}{{[}0.0, 0.1, 0.25, 0.5, 0.75, 0.9, 1.0{]}}                  &                   \\
Bayesian Ridge    & 'alpha\_init'   & \multicolumn{3}{l}{{[}1e-09, 0.0001, 1.0, 1000.0{]}}                              & lambda\_init'      & \multicolumn{3}{l}{{[}1e-09, 0.0001, 1.0, 1000.0{]}}                           &                   \\
Kernel Ridge      & 'kernel'        & {[}'linear', 'rbf', 'sigmoid'{]} & alpha'               & {[}0.001, 0.1, 1.0{]}   & 'gamma'            & \multicolumn{3}{l}{{[}1e-06, 0.001, 1.0, 1000.0{]}}                            &                   \\
K-Neighbors       & 'n\_neighbors'  & {[}3, 5, 8, 13{]}                & 'p'                  & \multicolumn{2}{l}{{[}1.5, 2.0, 3.0, 5.0{]}} &                                    &               &                           &                   \\
Support Vector    & 'kernel'        & {[}'linear', 'rbf', 'sigmoid'{]} & 'epsilon'            & {[}0.1, 1.0{]}          & 'C'                & {[}1e-06, 0.001, 1.0, 1000.0{]}    & gamma'        & \multicolumn{2}{l}{{[}1e-06, 0.001, 1.0{]}}   \\
Random Forest     & 'n\_estimators' & {[}30, 100, 200, 500{]}          & 'min\_samples\_leaf' & {[}1, 2, 3, 5{]}        & 'max\_features'    & \multicolumn{2}{l}{{[}0.3, 0.6, 1.0{]}}            &                           &                   \\
Gradient Boosting & 'n\_estimators' & {[}30, 100, 200, 500{]}          & 'min\_samples\_leaf' & {[}1, 2, 3, 5{]}        & 'learning\_rate'   & {[}0.01, 0.03, 0.1, 0.3{]}         & 'alpha'       & \multicolumn{2}{l}{{[}0.01, 0.1, 0.5, 0.9{]}}
\end{tabular}}
\caption{Parameters optimized through cross-validated gridsearch}
\end{table}

\begin{table}[]
\renewcommand{\thetable}{S\arabic{table}}
\begin{tabular}{lllllll}
                  & Length          & $\sigma$Length         & Diameter        & $\sigma$Diameter       & AR              & $\sigma$AR             \\
Elastic Net       & 21.1\%          & 39.5\%          & 26.1\%          & 56.5\%          & 17.2\%          & 16.3\%          \\
Bayesian Ridge    & 22.2\%          & 41.0\%          & 26.4\%          & 58.8\%          & 17.6\%          & 16.6\%          \\
Kernel Ridge      & 20.6\%          & 37.9\%          & 25.4\%          & 51.3\%          & 16.5\%          & 16.3\%          \\
K-Neighbors       & 22.0\%          & 33.2\%          & 21.6\%          & 39.5\%          & 12.4\%          & \textbf{13.6\%} \\
Support Vector    & \textbf{18.5\%} & 35.5\%          & \textbf{19.3\%}          & 43.0\%          & 14.2\%          & 15.0\%          \\
Random Forest     & 21.0\%          & \textbf{31.9\%} & 22.4\% & 38.6\%          & 11.1\%          & \textbf{14.0\%} \\
Gradient Boosting & 21.2\%          & \textbf{32.2\%} & 24.0\%          & \textbf{36.2\%} & \textbf{10.2\%} & \textbf{14.1\%}
\end{tabular}
\caption{Mean Absolute Percentage Validation Error for different regression algorithms fitting different parts of the dataset. All algorithms were optimized using gridsearchCV using the parameters and values given in table S1, results reported here are the errors from 5-fold cross validation of the best algorithm.}
\end{table}

\setcounter{figure}{0}
\renewcommand{\thefigure}{S\arabic{figure}}

\begin{figure}[htbp]
    \centering
    \includegraphics[width=\textwidth]{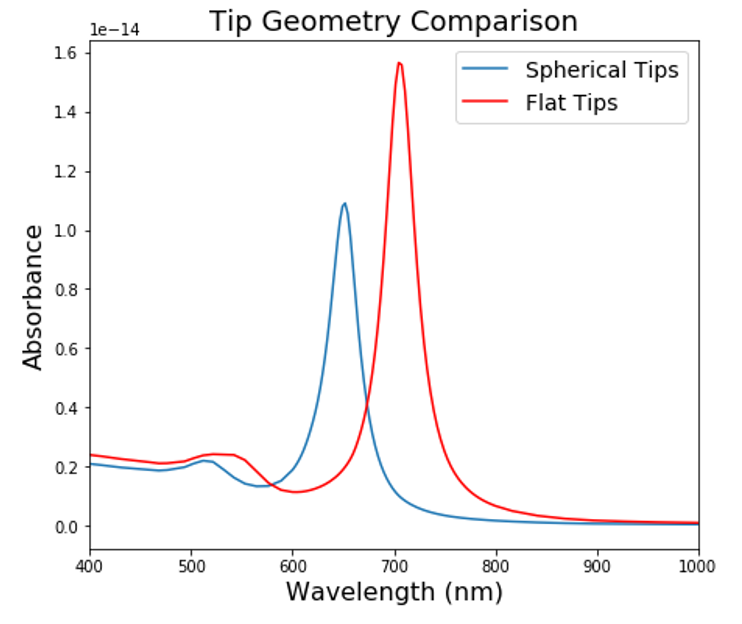}
    \caption{An illustration of the red shift caused by flat tipped AuNRs vs hemispherically capped AuNRs. The two spectra shown here are for rods of identical size, with the only differene being the tip geometry.}
    \label{figS1} 
\end{figure}

\begin{figure}[htbp]
    \centering
    \includegraphics[width=\textwidth]{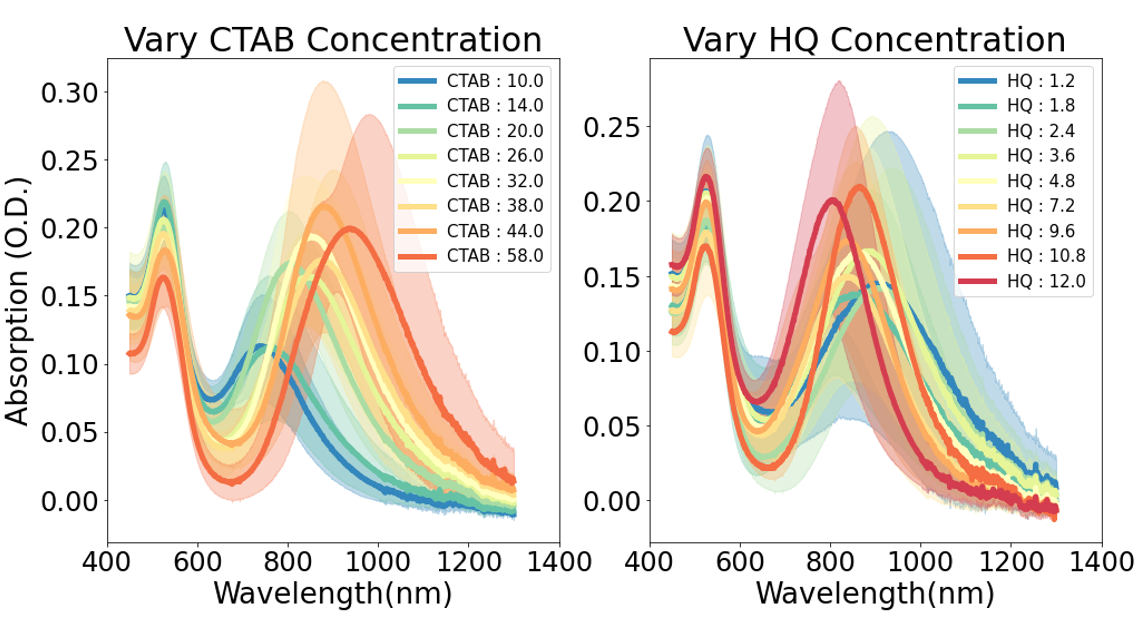}
    \caption{A visualization of a slice of the absorption spectra of the high-throughput synthesis dataset. All concentration values are given in mM. The left plot shows spectra of samples synthesized using varying concentrations of the surfactant, CTAB. The shadowed lines around the spectra show the standard deviations of the spectra, as these samples are averaged from 12 experiments using the same concentration of CTAB but different concentrations of the other reagents. This plot shows that there appears to be an inflection point between 14 and 20 mM where the samples go from making virtually no AuNRs to a moderately successful reaction (as shown by the increased intensity of the longitudinal peak relative to the transverse peak between the 14 and 20 mM spectra). Additionally, further increasing the CTAB concentration increases the yield and aspect ratios of the samples, although increasing the concentration to 58 mM appears to broaden the aspect ratio distribution significantly, as evidenced by the broadening of the longitudinal peak. The right plot shows a similar analysis of samples synthesized using varying concentrations of the weak reducing agent, hydroquinone (HQ). Increasing the concentration of HQ to 10.8 mM steadily increases AuNR yield and slightly blue shifts the longitudinal peak, but an increase past this concentration appears to begin to decrease yield, as the ratio of intensities of the longitudinal and transverse peaks decreases past this concentration.}
    \label{figS2} 
\end{figure}

\begin{figure}[htbp]
    \centering
    \includegraphics[width=\textwidth]{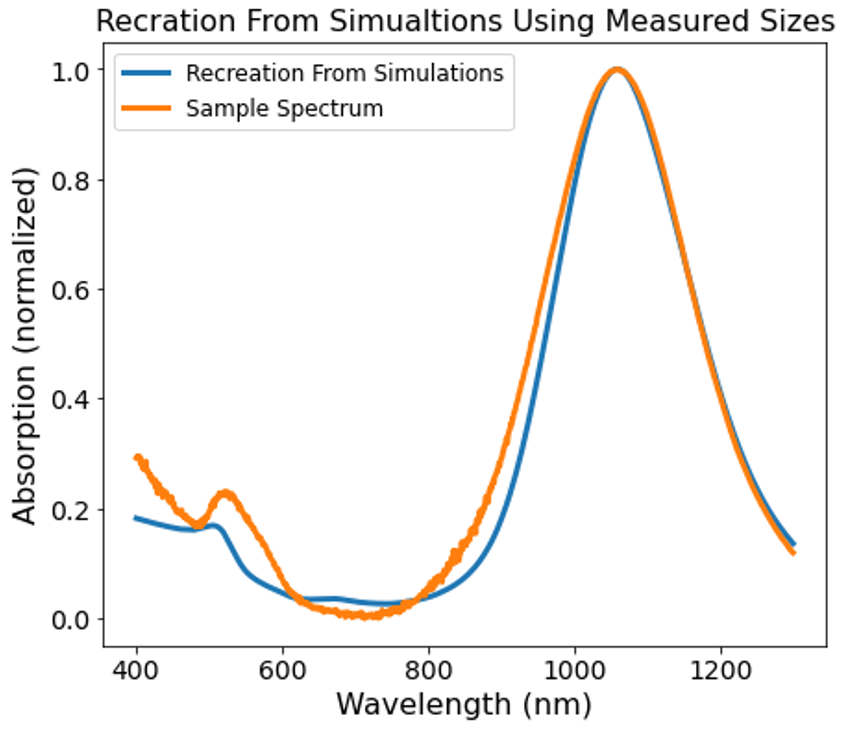}
    \caption{A comparison of an experimental spectrum and a simulated mixture spectrum recreated using the size distribution determined for this sample by directly measuring AuNRs from TEM images, showing a high degree of similarity between the experimental spectrum and the simulated mixture spectrum.}
    \label{figS3} 
\end{figure}

\begin{figure}[htbp]
    \centering
    \includegraphics[width=\textwidth]{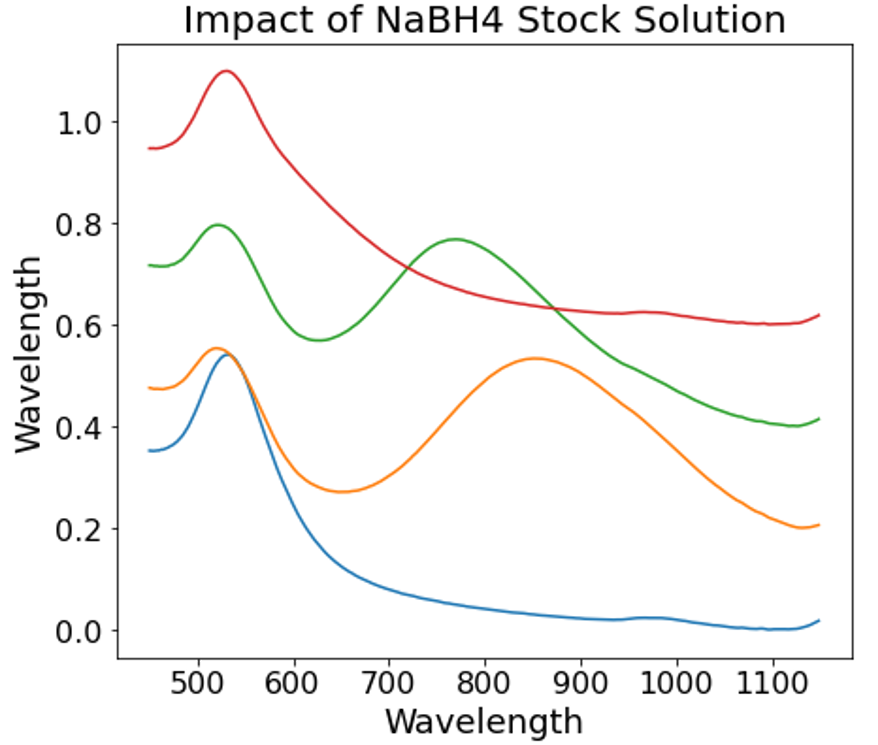}
    \caption{An illustration of the significant impact the NaBH\textsubscript{4} stock solution has on the resulting product. All 4 spectra have identical reaction conditions, the only difference being the NaBH\textsubscript{4} stock solution used. All four stock solutions are prepared to nominally identical concentration and purity. However, large differences appear to exist across the batches, with only two of the four (green and orange curves) successfully producing AuNRs.}
    \label{figS4} 
\end{figure}

\begin{figure}[htbp]
    \centering
    \includegraphics[width=\textwidth]{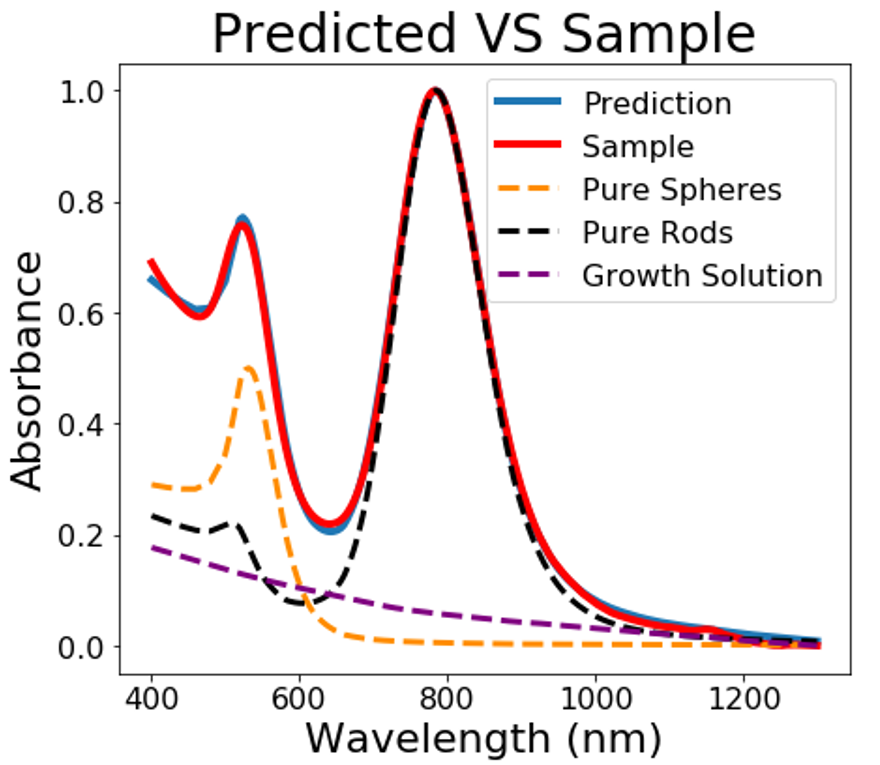}
    \caption{A recreation of a sample spectrum containing AuNS and unreacted growth solution using simulated AuNRs, AuNS, and a generated growth solution. This was validated by manual measurements of the AuNRs using TEMs to create the black curve, and then fitting the AuNS and growth solution fractions which produced the sample specturm. This shows how growth solution and spheres are needed to replicate the spectrum and how the growth solution especially influences the spectrum up to the NIR region.}
    \label{figS5} 
\end{figure}

\begin{figure}[htbp]
    \centering
    \includegraphics[width=\textwidth]{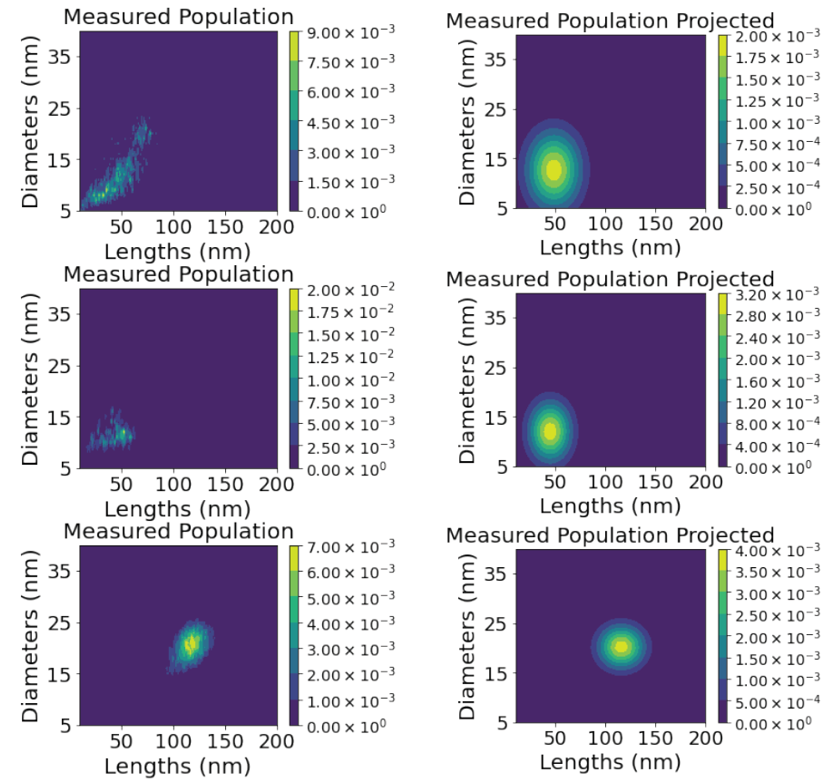}
    \caption{A few examples illustrating the projection of measured TEMs onto a normal distribution.}
    \label{figS6} 
\end{figure}

\begin{figure}[htbp]
    \centering
    \includegraphics[width=\textwidth]{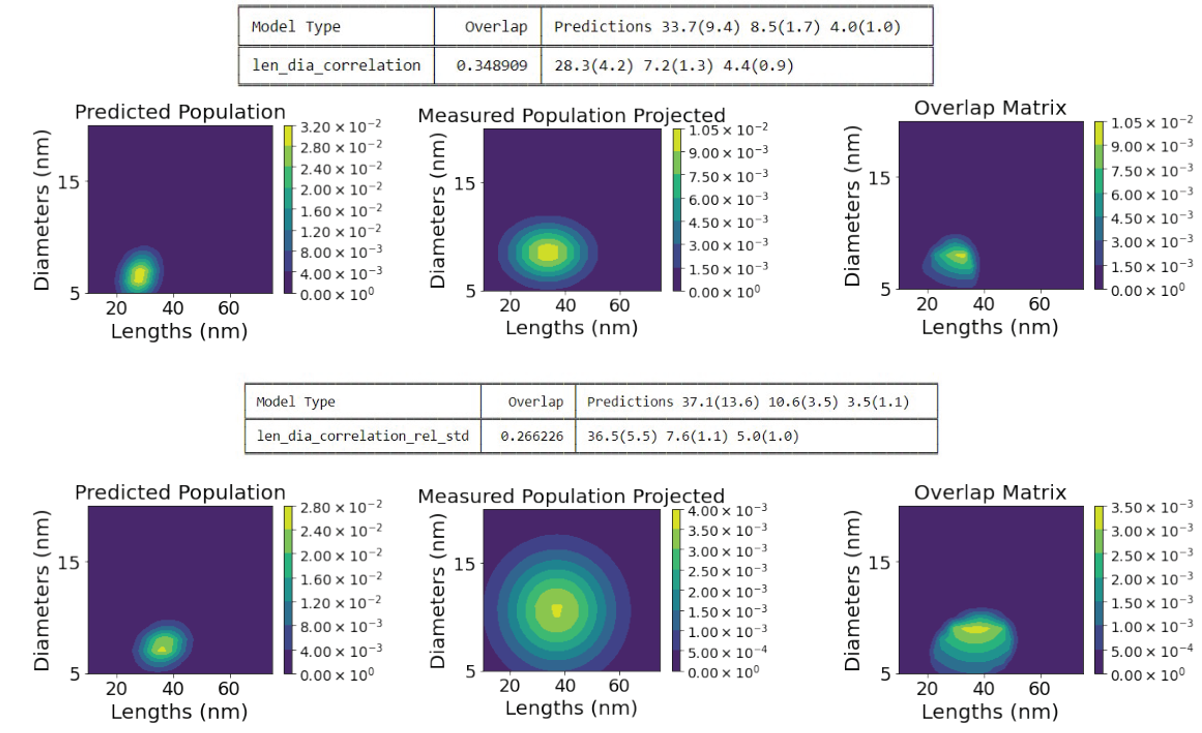}
    \caption{A few examples of samples with overlaps around the mean overlap, 0.3.}
    \label{figS7} 
\end{figure}

\begin{figure}[htbp]
    \centering
    \includegraphics[width=\textwidth]{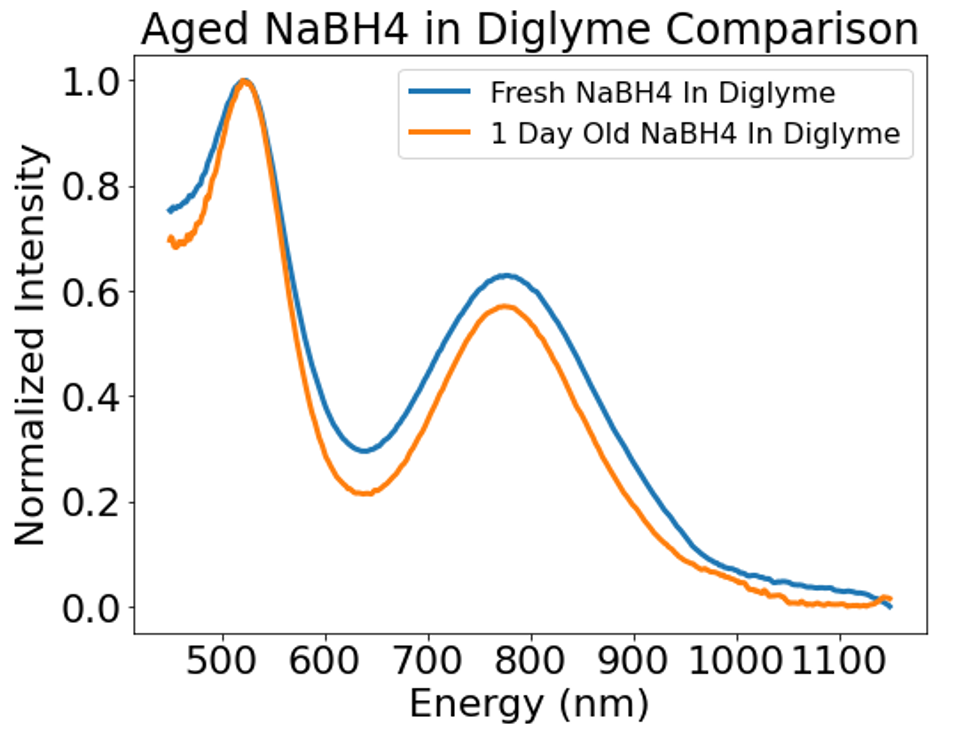}
    \caption{An example of repeat trials using the same synthesis condition with a stock solution of NaBH\textsubscript{4} in diglyme fresh and aged for 24 hours. The two conditions produce virtually identical spectra illustrating NaBH\textsubscript{4}'s stability in diglyme.}
    \label{figS8} 
\end{figure}

\begin{figure}[htbp]
    \centering
    \includegraphics[width=\textwidth]{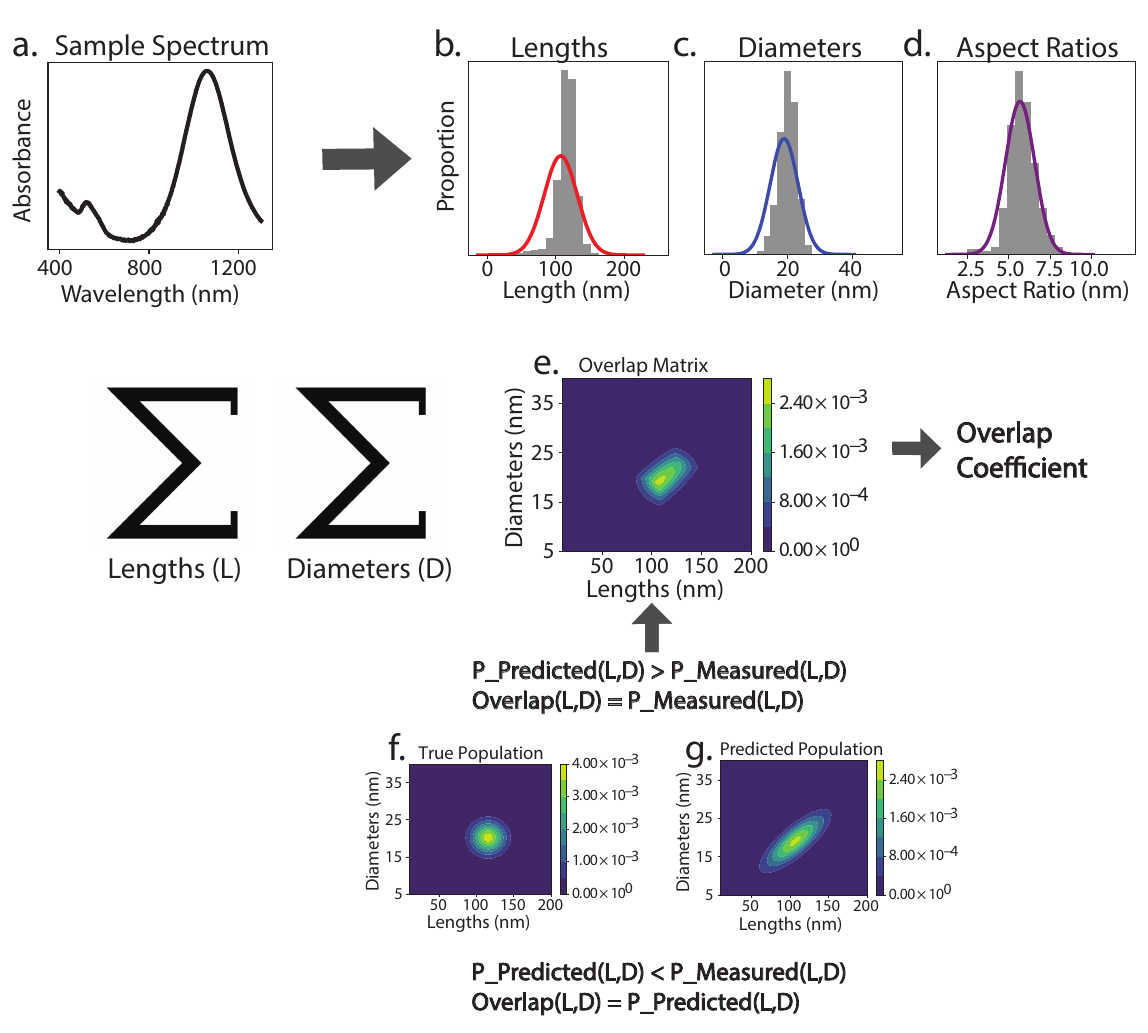}
    \caption{An illustration of how the accuracy of the spectral morphology analysis is represented with the overlap metric. The top row shows an AuNR spectrum (a) and the predicted length, diameter and aspect ratio distributions (b-d) on top of the corresponding measurements from TEM (gray histogram). To calculate the overlap between a predicted size distribution and the measured size distribution from TEM analysis (f,g), every point in the two distributions are compared to produce the overlap matrix (e), which is then summed to produce the overlap coefficient. This coefficient will be between zero and one, with a higher overlap indicating a more accurate prediction. More details on the overlap metric can be found in the SI, section "Overlap Metric"}
    \label{figS9} 
\end{figure}

\end{document}